# Carbon Dioxide as a Pollutant. The Risks of Rising Atmospheric CO₂ Levels on Human Health and on the Stability of the Biosphere.

Ugo Bardi,
Club of Rome and
Consorzio Interuniversitario di Tecnologia dei Materiali (INSTM)- Italy.
ugo.bardi@unifi.it

## Abstract

Nowadays, the increase of carbon dioxide concentration in the atmosphere is mainly discussed in relation to radiative forcing and the consequent global warming. However, $CO_2$ is a chemically active molecule that plays a vital role in Earth's ecosphere. $CO_2$ affects the acidity of seawater and has multiple effects on marine organisms. It is also a fundamental component of the photosynthesis and respiration reactions. There is evidence that higher $CO_2$ concentrations can make the photosynthetic reaction faster in some plants, but also negatively affect the respiration reaction in aerobic lifeforms. The effects of this chemical and biochemical perturbation on the biosphere and human health may be more important than generally highlighted in the discussion. These considerations stress the importance of rapidly reducing $CO_2$ emissions and, whenever possible, removing the excess from the atmosphere. They also show that geoengineering techniques based on Solar Radiation Management (SRM) alone cannot be sufficient to contrast the disruption caused by human $CO_2$ emissions.

## Introduction

The impact of increasing $CO_2$ concentrations in the atmosphere was discussed for the first time in 1896 by Svante Arrhenius in terms of its radiative forcing effect as an absorber and emitter of infrared radiation (IR)

[1]. Arrhenius didn't take into account the chemical and biochemical effects of $CO_2$, which started to be identified only about half a century later. The story is told in detail by Brewer [2], who reports that the first experiments on ocean acidity were carried out in the 1930s, but the term "ocean acidification" didn't become popular until the turn of the millennium [3]. Ocean acidification occurs when $CO_2$ dissolves in seawater to form carbonic acid ($H_2CO_3$), a weak acid that dissociates into bicarbonate ($HCO_3^-$) and hydrogen ions ($H^+$). Increased $H^+$ means increased acidity (lower pH). These high acidity levels have negative consequences, such as coral bleaching and damage to calcifying organisms. Nevertheless, ocean acidification is much less discussed than global warming and there is no agreed limit to ocean acidity equivalent to the internationally set temperature limit of the atmosphere.

The reactivity of $CO_2$ appears not just in seawater but in all the environments where $CO_2$ acts as an acid, either in aqueous environments where it generates hydrogen ions or when acting as a Lewis acid as an electron acceptor. Carbon dioxide is the keystone of the two main processes that create the planetary biosphere: photosynthesis and respiration. Without $CO_2$, there could be no life on Earth, but too much $CO_2$ is not necessarily a good thing.

The effect of $CO_2$ on the growth rate of some plants has been known for about a century [2]. This effect is measurable in terms of an increase in growth rates or leaf coverage, and it is considered the leading cause of the "global greening" phenomenon observed in recent years [4], [5]. Although $CO_2$ may increase agricultural yields, it is known that it does not generate an increase in the nutritional content of the food produced [6], [7].

On the other side of the biosphere cycle, respiration, the effects of high $CO_2$ concentrations are not easy to measure in quantitative terms but have been known since the 19th century under the name of "hypercapnia" (from the Greek *hyper,* "above" and *kapnos,* "smoke"). Common symptoms are dyspnea (breathlessness), nausea, headache, confusion, lethargy, and other symptoms. These effects are attributable to various factors but have been demonstrated to be related to reduced oxygen flow to tissues and to the brain [8]. It is known that $CO_2$ concentrations over ca. 50,000 ppm are lethal, while it is normally believed that values up to 5,000 ppm are acceptable for limited periods of time. Values under 1,000 ppm are considered safe inside homes. The effect of lower concentration is less clear, but recent results show that even lower concentrations can have measurable negative effects on the human metabolism, and affect the human brain in terms of the capability of performing complex tasks [9], [10], [11], [12], [13], [14], [15], [16], [17]. The results of these studies have been criticized for internal inconsistencies and other problems [18]. It is clear that we need more and better studies to determine with certainty the effect of $CO_2$ on human metabolism at these concentrations. But the available data nevertheless point to serious potential problems.

We are introducing into the environment an active substance that we know is lethal at high concentrations. We don't know what an acceptable lifetime exposure limit could be, and not even if it exists. The only thing we know is that current concentrations have never been experienced by human beings during their evolutionary history of the past few million years. Additionally, nowadays people tend to live in closed spaces where the $CO_2$ concentrations are typically higher than those in the open, not rarely well above 1000 ppm. The habit of wearing face masks in indoor environments has led to controversial assessments, but even though it may not be a critical problem, it can only increase the concentration of breathed $CO_2$ [19], [20]. The real problem, though, is that the $CO_2$ concentration in the atmosphere continues to rise at an increasingly faster rate, now being near 3 ppm per year. If this trend continues, it is clear that we are moving into an unknown territory with risks that cannot be neglected.

The present paper is an exploration of what's currently known about the metabolic effects of different atmospheric compositions and, in particular, of increasing $CO_2$ concentrations. It shows how the current atmospheric composition may have been an important factor in the development of the large brains, which are a characteristic of the "homo sapiens" species [21] and of other highly encephalized mammalian and bird species (e.g. dolphins, primates, ravens, and elephants) [22]. The current trends of increasing $CO_2$ concentrations may make the metabolic requirements of such large brains impossible to maintain.

The biochemical effects of $CO_2$ are also relevant to the current debate on geoengineering as Solar Radiation Management, SRM. In view of the projected biochemical damage of excessive $CO_2$ concentrations on the biosphere, radiation shielding makes sense only if coupled with actions aimed at controlling and reducing $CO_2$ concentrations.

## CO₂ in the Atmosphere

The early Earth's atmosphere is believed to have been composed mainly of $CO_2$ and to have contained no oxygen. With the evolution of oxygenic photosynthesis, maybe to 3.2-3.8 billion years ago, carbon dioxide started to be turned into organic carbon compounds and molecular oxygen [23]. The "Great Oxygenation Event" (GOE), around 2.3-2.4 Ga ago, started a trend of growth of atmospheric molecular oxygen that led to the appearance of aerobic organisms. The Phanerozoic Eon, starting 540 million years ago, saw oxygen concentrations rise to levels around and over 20% [25], [26], [27], remaining remarkably stable during the past 370 million years or so.

In parallel, $CO_2$ atmospheric concentrations showed a declining trend. During the Cenozoic Era, a robust trend of cooling and lowering $CO_2$ concentrations started about 55 million years ago [28]. During the last ca. 10 million years, the second half of the Miocene, the decline has been especially rapid [29]. The Pleistocene Epoch, the one that precedes the current one, the Holocene, saw $CO_2$ concentrations and temperatures oscillating together between ice ages and interglacial periods. During the ice ages, $CO_2$ concentrations reached values as low as 180 ppm, the lowest ever estimated during the Cenozoic, perhaps during the whole Earth's history.

These trends are best explained in terms of global ecosystem cycles that control $O_2$ and $CO_2$ concentrations in the atmosphere. Atmospheric oxygen is continuously created from $CO_2$ by photosynthesis in plants. It is removed by the respiration reaction in aerobic organisms and also by the "dark respiration" [30]. These reactions oxidize organic compounds at low temperature in the biosphere consuming oxygen and creating carbon dioxide. Oxygen is also removed at high temperatures by fires in the biosphere, and by the oxidation of "recalcitrant" carbon compounds stored in the crust [31], mainly kerogen, but also those compounds called "fossil fuels" by human beings. Recalcitrant compounds react massively with oxygen only at the very high temperatures that may be generated by massive volcanic eruptions of the type called "Large Igneous Provinces" (LIPs) [32] which are known to have generated substantial dips in oxygen concentrations [33] and mass extinctions during the Phanerozoic [34].

The $CO_2$ concentration in the atmosphere is regulated by geological and biological cycles. The main source of $CO_2$ is outgassing from the mantle [35], [36] which occurs at several locations, including mid-ocean ridges, rifts, volcanic arcs in subduction zones, orogenic belts, and large igneous provinces. Variations in the outgassing rate are related to different states of Earth's climate, known to have oscillated between extremes called "icehouses" and "hothouses." [37] These fluctuations are partly compensated by the feedback provided by the silicate weathering cycle, or "silicate reaction," which removes gaseous $CO_2$ from the atmosphere by reaction with the crust's silicates to form carbonates [38]. It is a very slow process from a human viewpoint, and it can be calculated that the $CO_2$ in the atmosphere is cycled in times of the order of hundreds of thousands of years [39] [40] [41]. This reaction involves a stabilizing feedback: it is faster with high temperatures, so it draws down $CO_2$, tending to cool the atmosphere [42]. The reaction rate is also affected by the activity of the biosphere, since plant roots tend to keep the soil humid, hence keeping silicates in contact with the $H^+$ ions generated by bicarbonate. A further factor that influences the $CO_2$ concentration is the gradual increase in solar irradiation over geological times, approximately 10% every billion years [43]. The feedback involved with the interplay of increasing irradiation and silicate erosion is believed to be the cause of the declining $CO_2$ concentration that has allowed the Earth system to maintain temperatures compatible with the presence of liquid water on the surface during the past 3-4 billion years.

The first to discuss the consequences of these trends on the biosphere were Lovelock and Whitfield in a 1982 paper [44]. They proposed that if the trend of CO2 decline were to continue for a long time, photosynthesis would eventually become impossible and life on Earth would disappear. They argued that the extinction of the biosphere would occur at a $CO_2$ concentration of about 150 ppm, and they calculated that, because of this effect, the lifespan of the biosphere is limited to no more than about 100 million years from now. However, Lovelock and Whitfield did not consider that the $C_3$ mechanism is not the only one used by vascular plants. If the "$C_4$" photosynthetic mechanism is taken into consideration, simple calculations [45], [46] led to the result that the biosphere could survive at least for nearly one billion years in the future. Recent calculations based on GCM climate models confirm these early results [47].

During the past few centuries, humankind has modified several vital ecosystem parameters, inverting these long-term trends. The combustion of fossil fuels and other factors are pushing $CO_2$ concentrations to values never observed during the past few million years, and the trend is projected to continue to even higher values in the coming decades, pushing Earth into a new hothouse status [48]. However. the consequences of this increase go beyond simply increasing planetary temperature but also involve chemical and biochemical effects. Life on Earth has been deeply affected by the varying $O_2/CO_2$ ratio in the atmosphere, a phenomenon described as the "battle of the gases" [33]. This subject will be discussed in depth in the following sections of the present paper.

## CO$_2$ and Photosynthesis

Carbon dioxide is one of the reactants of the oxygenic photosynthesis reaction powered by solar light. The overall reaction can be written as,

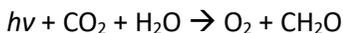

$hv + CO_2 + H_2O \rightarrow O_2 + CH_2O$

Here, the solar photon in the visible part of the spectrum is written as *hv,* and $CH_2O$ is the empirical formula for the glucose molecule ($C_6H_{12}O_6$). This way of writing the reaction is deceptively simple, but it implies intermediate reactants and catalysts active in a series of steps occurring inside specific cells called "chloroplasts."

The reaction occurs in two stages: the first is the photocatalytic reaction, which splits water using as a catalyst a molecule called "chlorophyll." The second (called "dark phase" or "dark stage" or the "Calvin Cycle") decomposes $CO_2$ and makes it react with hydrogen atoms to create organic compounds. It uses as a catalyst an enzyme called "ribulose bisphosphate carboxylase/oxygenase," or "Rubisco." The energy source for this part of the cycle is adenosine triphosphate (ATP), created in different cells by the respiration process.

For unicellular organisms, the reactants of the photosynthesis reaction can be transported to and from the reaction centers simply by moving through the lipid bilayer of the cell membrane. For vascular plants, the exchange requires specific channels called "stomata" that connect the cells to the atmosphere. This connection is direct in the case of the "C$_3$" photosynthesis mechanism, the ecosystem's oldest and most common one. However, in the current $CO_2$-poor conditions, the reverse process "dark respiration" or "photorespiration" affects the reaction rate. Rubisco has an affinity with oxygen, and when solar light is not powering the reaction toward creating organic compounds, it produces $CO_2$ by oxidizing organic compounds.

The $C_4$ photosynthetic mechanism [49], [50] appeared in vascular plants around 10 million years ago, at the start of the rapid fall in the $CO_2$ concentrations of the mid-Miocene [29]. Today, it accounts for only about 3% of plant species but contributes around 25% of global terrestrial photosynthesis. It uses the same molecular machinery as the older $C_3$ mechanism, but the reaction center where the rubisco action occurs is no longer directly connected to the atmosphere. Instead, $CO_2$ is transformed into malic acid by the phosphoenolpyruvate (PEP) carboxylase enzyme and then accumulated in "bundle sheath" cells. It is later converted into $CO_2$ by specific enzymes and then transferred to the Rubisco reaction center. Succulent plants (Crassulaceae and Cacti) use a third mechanism, CAM (crassulacean acid metabolism), which makes them even more resistant to arid conditions. The $C_4$ reaction path requires more energy than the $C_3$ one, but it is faster and more efficient for low $CO_2$ concentration. Since the stomata remain closed longer, $C_4$ plants do not transpire large amounts of water and are more resistant to arid conditions [51].

The $C_4$ mechanism was never adopted by trees, except for some species of Euphorbiae, not taller than ca. 10 m. [52]. That may be related to how trees use the transpiration mechanism to create the pressure difference that pulls water and dissolved minerals from the roots to the leaves through the xylem [53], which cannot function efficiently with the $C_4$ mechanism, a point not often highlighted in the scientific literature on the subject. Transpiration from trees, coupled with evaporation from the soil ("evapotranspiration"), also has substantial effects on the hydrological cycle, influencing the moisture regime [54]. $C_4$ plants also cannot generate the transport of water from the oceans inland that trees can accomplish using the "biotic pump" [55]. Because of these factors, forests are poorly adapted to low-$CO_2$ environments. Indeed, during the last glacial maximum, about 20,000 years ago and with a $CO_2$ concentration as low as 180 ppm, forests were reduced to sparse patches surrounded by steppes or savannas [56]. However, other factors may also have been active, such as low temperatures, effects of the megafauna, and hominins using fire to clear forests [57], [58].

The current epoch, the Holocene, started about 12,000 years ago and saw a reversal of the Pleistocene trends. Increasing temperatures and $CO_2$ concentrations led to a relatively stable climate that has lasted up to recent times. Anthropogenic emissions of greenhouse gases in agriculture may have been responsible for this effect [59], but the industrial age changed everything. The effect of anthropogenic $CO_2$ emissions on temperatures is well known and documented, but biochemical effects on photosynthesis are also starting to be noticeable. One is the global increase in forest cover called "global greening" [4], [60], [61]. Not all observations are consistent with the extent of this new greening, and some data indicate that it may be accompanied by "global browning"[62], [63]. But it seems certain that it is due in large part to $CO_2$ fertilization [64].

The effect of increased atmospheric $CO_2$ on crop production is a more complex matter, where several incompatible claims have been made. At least one source proposed that all the increased crop yields in the past 70 years or so (the so-called "green revolution") are due to $CO_2$ fertilization [65]. Other studies are more cautious, and the results of the "Free Air $CO_2$ Enrichment (FACE)" [66] experiments found that the main factors affecting crop yield are irrigation and temperature. Zheng [67] found that productivity tends to taper off and then decline for $CO_2$ concentration over around 1000 ppm. That's not surprising since vascular plants are aerobic organisms, and, as we'll see in the next section, high $CO_2$ concentrations damage their metabolism. $C_4$ plants (maize, millet, sugarcane) show little or no increase in productivity for increasing $CO_2$ concentrations [68]. The crucial point, in any case, is that the increased yield in total biomass produced is not accompanied by a corresponding increase in the nutritional content of the food [6].

## $CO_2$ and Respiration.

The respiration reaction in aerobic organisms is called the "Krebs Cycle" or "Citric Acid Cycle." It takes place in specialized cells called "mitochondria." It can be schematically written as follows, with "ATP" standing for adenosine triphosphate, the "fuel" for most metabolic reactions in living beings:

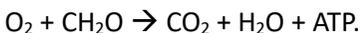

$O_2 + CH_2O \rightarrow CO_2 + H_2O + ATP.$

This reaction looks nearly identical to the photosynthesis one, except that it goes in the opposite direction. Yet, the molecular mechanisms involved are completely different. Note that the Krebs cycle is thermodynamically favored, so the reaction is not reversible, unlike the Calvin cycle. Note also that in aerobic organisms, oxygen is not just involved in the synthesis of ATP, but it intervenes directly in a large number of metabolic reactions [69].

To provide the reactants and remove the products of the respiration reaction, unicellular aerobic organisms can simply exchange oxygen and carbon dioxide by diffusion through the cell's lipid bilayer membrane. For multicellular organisms, this mechanism cannot function, and gases must be transported inside and outside the body. Small metazoa, such as insects, transport $O_2$ and $CO_2$ in gaseous form through an open tracheal system. Larger metazoa transport gases dissolved in a water-based liquid (blood). The exchange with the atmosphere is carried out at the interface of the pores of specific organs (gills or lungs). Then, blood is pumped everywhere in the body using a positive displacement pump (heart).

$CO_2$ is soluble in water as bicarbonate ions ($HCO_3^-$); hence, it can be directly dissolved in the aqueous part of the blood, the plasma. The reaction is helped by a specific enzyme [70]. $O_2$, instead, is a non-polar molecule not easily dissolved in polar liquids. For this reason, it is transported in blood by special molecules contained inside specialized cells called "red blood cells" or "erythrocytes." In mammals, hemoglobin is the transporting molecule: it can bind up to four oxygen molecules and transports approximately 97% of the oxygen in arterial blood. Some metazoa use other compounds, but they all play the same role. Hemoglobin can also transport carbon dioxide bound to it to form the compound called "carbaminohemoglobin." A significant fraction, about 10%, of the $CO_2$ present in blood processes is transported in this way in the mammalian body.

One point not often discussed in the literature is that $CO_2$ is not just an inert product of the respiration reaction but also affects the oxygen transport rate. Indeed, people who suffocate in closed environments (e.g., submariners or miners) do not die for lack of oxygen, but for the poisoning effect of increased $CO_2$, which affects oxygen transport in the body. For the $O_2/CO_2$ exchange mechanism to work, hemoglobin must act as a "truck." It must load oxygen at the alveoli, in the lungs, and unload it near the cells that need it. It is called the "Bohr effect" when referring to oxygen binding/unbinding and the "Haldane effect" when referring to the parallel and opposite binding/unbinding of $CO_2$ [71].

How does hemoglobin "know" when it is the right moment to do one or the other thing? It is a typical example of "allosteric regulation," meaning that a change in the molecule's properties is generated by a conformational change. Hemoglobin can assume two different allosteric states: the "R-state" (relaxed state), which makes it release $O_2$, and the "T-state" (tense state), which causes it to bind $O_2$. The $CO_2$ molecule acts as an allosteric regulator. When it is dissolved in the blood as bicarbonate the resulting acidic environment stabilizes the T state of hemoglobin, promoting the release of oxygen. In addition, the N-terminal amino groups of hemoglobin's α-subunits and the C-terminal histidine of the β-subunits can become protonated in acidic conditions. This protonation enhances ionic interactions that stabilize the T state, further facilitating oxygen unloading. Finally, $CO_2$ can also react with the N-terminal amino groups to form carbamates, which further stabilizes the T state and contributes to the release of oxygen. This reaction generates additional protons, reinforcing the acidic environment and promoting the Bohr effect.

The key point of these effects is that the concentrations of $CO_2$ and of $O_2$ in the blood affect each other. More $CO_2$ in the atmosphere means less oxygen is transported in the blood, and that must affect the metabolic rate of metazoa. This effect can be experimentally detected on the human brain at relatively large $CO_2$ concentrations [8] but it is probably active at much lower concentrations.

Excess $CO_2$ in the atmosphere also directly affects metabolism in other ways. The references cited in the introduction [9-17] discuss several negative effects. Without going into the details, increased $CO_2$ content can reduce the blood's pH. The body reacts to contrast this effect by mobilizing $Ca^{2+}$ ions from bone tissue to replace $H^+$ ions. The effect may be the calcification of organs such as kidneys and arterial walls and affecting the neuron activity in the brain. This and other effects may generate respiratory failure, cardiac diseases, cognitive impairment, and more. Other effects attributed to high $CO_2$ concentrations are metabolic syndrome [14], type 2 diabetes, obesity, sleep disorders, and more.

$CO_2$ concentrations between 600 and 1000 ppm have been reported to be already sufficient to generate measurable reductions in cognitive abilities [18]. These results should be considered as still preliminary, but they need to be taken into account in view of the modern tendency for humans to live indoors in scarcely ventilated spaces where the $CO_2$ concentration may be at least double that of the open air. This effect might explain the reversal of the trend of increasing human intelligence reported for the first time by Flynn in 1984 [72]. The "reverse Flynn effect" (aka "global dumbing"), has been recently detected [70], and it has been attributed to environmental factors [71]. It cannot be excluded that these environmental factors include the current increase in $CO_2$ atmospheric concentrations [15].

# Discussion

Carbon dioxide emissions are not the only perturbation of the ecosphere caused by human activities. Other factors involve other greenhouse gases (e.g., $CH_4$), deforestation, particulate emission, biodiversity loss, disruption of the hydrological cycles, albedo changes, and more. Nevertheless, $CO_2$ remains a keystone element of the ecosphere's temperature regulation and metabolism, having played this role during the past 3-4 billion years. So, we cannot ignore that varying its concentration will have important effects on Earth's biosphere. These effects are not just to be understood in terms of temperature forcing, but also in terms of chemical and biochemical perturbations of the biosphere, including ocean acidification and alteration of the metabolism of aerobic organisms.

The direct negative effects of increasing $CO_2$ concentrations in the atmosphere are often neglected on the basis of highly simplified arguments. A typical one is that $CO_2$ concentrations have been much higher in the past so there is nothing to be worried about. It is true that the biosphere can survive the increase because it did exist at concentrations several times higher than the current ones during the first part of the Cenozoic [28]. But today's biosphere is not the same that existed during the early Cenozoic and we have strictly no data about how the metabolism of the lifeforms of those ancient times were adapted to these high $CO_2$ levels. Pushing the current biosphere back to the conditions of those ancient times may have unexpected and unfavorable consequences on the metabolic processes of an ecosystem that had adapted to the present conditions over several millions of years. In particular, we cannot exclude that the evolution of large brains in hominin species during the Pleistocene was made possible by the high oxygenation of the tissues generated by the high $O_2/CO_2$ ratio, [73] [33].

Another weak argument is that high concentrations of $CO_2$ are a good thing because $CO_2$ is "food for plants" [74]. It is true that many plants grow faster in the presence of higher concentrations of $CO_2$, but the argument is self-defeating since it confirms that even small changes in the $CO_2$ concentration in the atmosphere can have large effects on the biosphere. These effects are not necessarily good, especially for those organisms that are not plants (e.g. human beings). Additionally, the global greening effect has not been able to reverse or slow down global warming up to now.

Finally, the faster plant growth does translate into a larger production of food for humans, since the amount of nutrients contained in $CO_2$-grown crops remains the same as in ordinary kinds of cultivation [6], [7]. It has been said (Lemon 1977, as cited in [2]) that these products are a way of *"selling more water to the housewife packaged in green leaves."*

Even without agreeing with the several indications that high $CO_2$ concentrations are bad for human health and for the whole biosphere, it is generally accepted that when dealing with a substance with potentially negative effects on human health, producers and users carry the burden of proof that these effects do not exist or that can be controlled by setting exposure limits. In the case of $CO_2$ injected into the atmosphere, we are subjecting humankind to exposure to a substance known to be lethal at high concentrations and for which we have no evidence for an exposure limit that would pose no risk to people and living beings. Even if the current $CO_2$ concentrations could be demonstrated to pose no long-term risks, we keep increasing the exposure at a rate which is, at present, of ca. 3 ppm per year, and that may well increase in the future since complex systems, such as Earth's atmosphere, typically show non-linear behavior.

## Conclusion

The data reported in the present paper highlight the vital importance of reducing and eliminating CO2 emissions as fast as possible, not just in view of global warming, but taking into account the negative chemical and biochemical effects of CO2. At present, it is not possible to determine upper limits for $CO_2$ concentrations that could be considered "safe" for human health. However, it is likely that it cannot be much higher than the average Holocene one (280 ppm), before the industrial era, considering that *Homo sapiens* evolved in an environment with concentrations typically below 320 ppm during the Pleistocene and potentially up to 400 ppm during the Pliocene. The current levels of $CO_2$ in the atmosphere, exceeding 420 ppm, are the highest in human history and far exceed the atmospheric conditions that early humans and their ancestors experienced. In the future, restoring atmospheric $CO_2$ concentrations at levels close to pre-industrial ones may be a vital task for the survival of humankind as we know it. Carbon capture and sequestration (CCS) [73], [74] may be an important tool for this purpose, but the restoration of degraded planetary ecosystems may be just as important and less expensive in the short run [75], [76].

Disclaimers: the author declares no conflict of interest. This work was not supported by any funding agencies. No parts of the text were written by AI programs.